\documentclass[conference,letterpaper]{IEEEtran}
\IEEEoverridecommandlockouts

\usepackage{mathpazo}
\usepackage{times}
\usepackage{amsmath}
\usepackage{amsfonts}
\usepackage{latexsym}
\usepackage{amssymb}
\usepackage{mathabx}
\usepackage{float}
\usepackage{upref}
\usepackage{theorem}
\usepackage{graphicx}
\usepackage{psfrag}
\usepackage{cite}

\usepackage{balance} 
\usepackage{url}

\usepackage{color}

\usepackage{tikz}
\usetikzlibrary{arrows.meta,bending}

\makeatletter
\newcommand\curvearrowed@[3]
  {%
    \begin{tikzpicture}
      \node[inner sep=.05ex](a){\kern-.25ex#3\kern-.05ex};
      \draw[arrows={-Latex[#1]},line width=#2]
        (a.north west) to[bend left=45] (a.north east);
    \end{tikzpicture}%
  }
\newcommand\curvearrowed[1]
  {%
    \relax
    \ifmmode
      \mathchoice
        {\curvearrowed@{}{.4pt}{$\displaystyle #1$}}
        {\curvearrowed@{}{.4pt}{$\textstyle #1$}}
        {\curvearrowed@{scale=.8}{.325pt}{$\scriptstyle #1$}}
        {\curvearrowed@{scale=.7}{.25pt}{$\scriptscriptstyle #1$}}%
    \else
      \curvearrowed@{}{.4pt}{#1}%
    \fi
  }
\makeatother

\usepackage{algorithm,algpseudocode}

\makeatletter
\newcommand{\removelatexerror}{\let\@latex@error\@gobble}
\makeatletter
\newcommand{\proofpart}[2]{%
	\par
	\addvspace{\medskipamount}%
	\noindent\emph{Part #1: #2}\par\nobreak
	\addvspace{\smallskipamount}%
	\@afterheading
}
\makeatother

\theoremstyle{plain}
\theorembodyfont{\normalfont\slshape}

\newtheorem{thm}{Theorem$\!$}
\newenvironment{theorem}
{\begin{thm}\hspace*{-1ex}{\bf.}}{\end{thm}}

\newtheorem{clm}[thm]{Claim$\!$}
\newenvironment{claim}{\begin{clm}\hspace*{-1ex}{\bf.}}{\end{clm}}

\newtheorem{lem}[thm]{Lemma$\!$}
\newenvironment{lemma}{\begin{lem}\hspace*{-1ex}{\bf.}}{\end{lem}}

\newtheorem{prop}[thm]{Proposition$\!$}

\newtheorem{cor}[thm]{Corollary$\!$}
\newenvironment{corollary}{\begin{cor}\hspace*{-1ex}{\bf.}}{\end{cor}}

\newtheorem{defn}[thm]{Definition$\!$}

\newtheorem{xmpl}[thm]{Example$\!$}

\newtheorem{cnstr}{Construction$\!$}

\newtheorem{rmk}[thm]{Remark$\!$}

\newcounter{enumrom}
\renewcommand{\theenumrom}{(\roman{enumrom})}

\makeatletter
\renewcommand{\@endtheorem}{\endtrivlist}
\makeatother

\makeatletter
\renewcommand{\thefigure}{{\@arabic\c@figure}}
\renewcommand{\fnum@figure}{{\bf Figure\,\thefigure}}
\makeatother

\newcommand{\cC}{\mathcal{C}}

\newcommand{\cG}{\mathcal{G}}

\newcommand{\cO}{\mathcal{O}}
\newcommand{\cP}{\mathcal{P}}

\newcommand{\cT}{\mathcal{T}}

\newcommand{\bs}{\mathbf{s}}
\newcommand{\bt}{\mathbf{t}}

\newcommand{\bx}{\mathbf{x}}
\newcommand{\by}{\mathbf{y}}
\newcommand{\bz}{\mathbf{z}}

\newcommand{\bfs}{{\boldsymbol s}}
\newcommand{\bft}{{\boldsymbol t}}

\newcommand{\bfx}{{\boldsymbol x}}

\renewcommand{\leq}{\leqslant}

\renewcommand{\geq}{\geqslant}

\newcommand{\Cref}[1]{Co\-ro\-lla\-ry\,\ref{#1}}

\outer\def\proclaim #1. #2\par{\medbreak
 \noindent{\bf#1.\enspace}{\sl#2\par}%
 \ifdim\lastskip<\medskipamount \removelastskip\penalty55\medskip\fi}

\begin{document}

\title{\textbf{Semiquantitative Group Testing \\ in at Most Two Rounds}}

%

\author{
   \IEEEauthorblockN{Mahdi Cheraghchi}
   \IEEEauthorblockA{Department of EECS\\
     University of Michigan, Ann Arbor, MI\\
     Email: {mahdich@umich.edu}\thanks{M.\ Cheraghchi's research was partially supported by the National Science Foundation under Grant No.\ CCF-2006455.}} 
\and
   \IEEEauthorblockN{Ryan Gabrys}
   \IEEEauthorblockA{Naval Information Warfare Center, \\
   San Diego, and UIUC, Urbana, IL \\
     Email: {ryan.gabrys@gmail.com}} 
\and
   \IEEEauthorblockN{Olgica Milenkovic}
   \IEEEauthorblockA{Department of ECE\\
     University of Illinois, Urbana, IL\\
     Email: {milenkov@illinois.edu}} 
 }

\maketitle
\begin{abstract} Semiquantitative group testing (SQGT) is a pooling method in which 
the test outcomes represent bounded intervals for the number of defectives. 
Alternatively, it may be viewed as an adder channel with quantized outputs. SQGT represents a
natural choice for Covid-19 group testing as it allows for a straightforward interpretation of the
cycle threshold values produced by polymerase chain reactions (PCR). Prior work on SQGT did not
address the need for adaptive testing with a small number of rounds as required in practice.
We propose conceptually simple methods for $2$-round and nonadaptive SQGT that significantly improve upon existing 
schemes by using ideas on nonbinary measurement matrices based on expander graphs and 
list-disjunct matrices.
\end{abstract}

\section{Introduction}

Group testing (GT) is a scheme designed to efficiently identify a small set of subjects with a particular property (standardly referred to as defectives) within a large population, first introduced by Dorfman~\cite{D43} and further studied in many other works, including~\cite{KS64,DH06,kairouz}. Group testing entails testing a collection of carefully selected subpopulations and reporting for each subgroup a binary answer: A positive answer is indicative of the existence of at least one defective in the subgroup while a negative answer implies the absence of defectives. Given that screening protocols are extensively used in engineering and science, group testing has found wide-spread applications in communication theory, signal processing, computer science, and computational biology~\cite{W85,DH06}.

Many different variants of group testing have been proposed in the literature~\cite{DH06,D43,D04}. These include threshold group testing proposed by Damaschke~\cite{D06} and quantitative (additive) group testing studied by Lindstr\'om and Du and Hwang~\cite{L75,DH00,D04}. In the latter case, the test results report the exact number of defectives in the test subpool. In the former case, if the number of defectives in a test is smaller than a lower threshold, the test outcome is negative; if the number of defectives is larger than an upper threshold, the test outcome is positive; otherwise, the result is arbitrary (positive or negative). To bridge the two above described paradigms, Emad and Milenkovic~\cite{AM110,EO14,EO16} introduced the notion of semiquantitative group testing (SQGT). SQGT represents a unifying framework of a number of testing protocols, including conventional, quantitative and gapless threshold group testing and the schemes by D'yachkov and Rykov~\cite{DR81,DR83}. In SQGT, the result of a test is a nonbinary value that depends on the number of defectives through a fixed set of thresholds. The SQGT model may also be viewed as a quantitative group testing method followed by a quantizer. The original motivation for introducing SQGT models is genotyping; more recently, the model has been used by Gabrys et al.~\cite{acdc} to describe the test outcomes of a Covid-19 testing process known as real-time reverse-transcriptase polymerase chain reaction (PCR).  

In \emph{nonadaptive} SQGT, each subject is assigned a unique binary or nonbinary indicator word of length equal to the total number of tests. These indicators are arranged column-wise in a \emph{test matrix}. Each coordinate in the codeword assigned to a subject corresponds to a test, and its value reflects the ``concentration'' of the sample corresponding to the given subject in the test. Note that the concentrations are nonnegative integers that usually correspond to the number of units of the genetic material of an individual subject. Two families of nonadaptive SQGT codes, SQ-disjunct and SQ-separable, were analyzed in~\cite{EO14,EO16}. In the same work, a number of constructions for nonadaptive uniform and nonuniform (quantized) SQGT codes were presented but no results were reported for adaptive tests. The more recent work~\cite{acdc} introduced the first combinatorial and probabilistic adaptive SQGT (ASQGT) schemes, the former extending the work of Hwang~\cite{gbgt75} on generalized binary group testing. The proposed combinatorial ASQGT schemes involve what is referred to as parallel and deep search methods that lead to a relatively large number of testing rounds. This is an undesirable feature for practical implementations of SQGT in Covid-19 testing.

Here, we describe the first known combinatorial two-round adaptive SQGT (ASQGT) for a special selection of (quantization) thresholds studied in~\cite{acdc}. The scheme uses $O\Big(\frac{d \log\log \tau}{\log \tau} \log \frac{n}{d} \Big )$ tests for $n$ subjects, $d$ defectives and $\tau$ SQGT thresholds. It builds upon the ideas of list-disjunct group testing~\cite{list-disjunct} and like the approach~\cite{acdc} uses nonbinary test matrices obtained by careful linear combining of the rows of a binary disjunct matrix. The described two-round ASQGT protocol differs from the information-theoretic bound only by about a factor $\log \tau$. 
We then proceed to improve existing nonadaptive protocols by extending the construction of Porat and Rothschild~\cite{pr11}.

The paper is organized as follows. Sections~\ref{sec:tworounds} describes our main result, the first known two-round ASQGT. Section~\ref{sec:oneround} presents new nonadaptive SQGT schemes that significantly improve upon previous constructions~\cite{EO14,EO16} and imply new upper bounds for nonadaptive SQGT. 


\section{Terminology, GT Background, and Bounds}\label{sec:back}

We start with some relevant terminology. All parameters are denoted by small-case letters, while vectors and matrices are denoted by  bold-face small-case and capitalized Latin letters, respectively. Entries of the vectors are indexed by subscripts while matrix entries are indexed by pairs of integers within parentheses. Unless stated otherwise, all $\log$s are to base $2$.

Assume that there are $n>1$ test subjects labeled by elements in $[n]:=\{1,\ldots,n\}$ among which $d<n$ are defective (i.e, infected). 
In conventional group testing, we summarize the set of tests through a binary matrix ${\bf B}^{m \times n}$ in which every column of the matrix uniquely characterizes an individual and each row represents a test. The $(i,j)^{\text{th}}$ entry of $\bf B$, $\bf{B}$$(i,j),$ equals~$1$ if and only if the individual labeled $j$ is included in the $i^{\text{th}}$ test. Let $\bt_{I} \in \{0,1\}^m$ denote the binary vector that results from $m$ tests using $\bf B$, assuming that the set of infected individuals equals $I \subset [n]$, with $|I|\leq d$. Whenever clear from the context, we omit the subscript $I$. In conventional group testing $\bt_{I}(l)=1$ if and only if the $l^{\text{th}}$ test includes at least one element from $I$. Let $\bt_{L} \in \{0,1\}^m$ be defined analogously for another set $L \subset [n]$. We say that a set $L$ is consistent with $I$ if $\bt_{L} \leq \bt_{I}$ entrywise. 

The matrix ${\bf B}^{m \times n}$ is termed $d$-disjunct if no vector $\bt_{I}$ for $|I|\leq d$ contains in its support a column of  $\bf B$ not indexed by $I$. The disjunctness property ensures that the test results obtained from $B$ uniquely identify the set of defectives. A matrix $\bf B$ is termed $(d,\ell)$-list-disjunct if the tests output a superset of the defectives of size at most {$\ell+d$}; for such a matrix, the size of any list $L$ consistent with $I$ is at most {$\ell+d$}. Clearly, a matrix $\bf B$ which is $d$-disjunct is equivalent to one which is {$(d,0)$}-list-disjunct.
The notion of list-disjunct matrices was explicitly formulated (in an equivalent form) in \cite{INR10} and is also essentially equivalent to what was defined earlier in \cite{C09}.

We review the following known results pertaining to the existence of $(d,\ell)$-list-disjunct test matrices ${\bf B} \in \{0,1\}^{m \times n}$ with $\ell = \cO(d)$ and $m = \cO(d \log \frac{n}{d})$. First, note that it is straightforward to see that for a maximal $L$ one has $I \subseteq L$. Therefore, as noted in~\cite{BGV05}, the existence of a $(d,\ell)$-list-disjunct test matrix ${\bf B} \in \{0,1\}^{m \times n}$ with $\ell = \cO(d)$ naturally implies a two-round testing scheme: The first round of tests is governed by the rows of ${\bf B}$ while the second round involves individually testing subjects in $L$. 
Randomized and explicit constructions of list-disjunct matrices exist, particularly via expander graphs \cite{BGV05,DR83,list-disjunct,C09,INR10}. The best known construction which achieves an optimal number of rows and nearly linear time recovery (in the number of rows) is given by \cite{ref:CN20}. 

The best lower bound on the number of tests necessary for an adaptive ASQGT scheme was established in~\cite{acdc} via a simple counting argument and the bound equals $\frac{d}{\log \tau} \log \frac{n}{d}$. In the next section, we establish the existence of a two-round scheme that differs from this lower bound by a factor of $\log \log \tau$ only. For the single-round setting, using a variation of the argument employed by F{\"u}redi \cite{F96} in the context of cover-free codes, one can show that the corresponding number of tests scales as $\frac{d^2}{(\log \tau)^3} \log_d n$ whereas the construction from Section~\ref{sec:oneround} implies the existence of a scheme that requires at most $\frac{d^2}{\log \tau} \log n$ tests. This lower bound applies to not only general nonadaptive SQGT, but in fact the particular \emph{saturation model} as well, which is the focus of this work. The derivation of the bound is relegated to the full version of the paper.

\section{Two-Round ASQGT} \label{sec:tworounds}

Let $\cG$ be a bipartite graph with a vertex partition $\cP$ (people) and $\cT$ (tests) such that every vertex in $\cP$ has degree $k$ (i.e., $k$ neighbors) and $|\cP|=n$, $|\cT|=m$. We say that $\cG$ is an $(\alpha, \beta)$-expander if every $P \subseteq \cP$ of size at most $\alpha |\cP|$ has at least $\beta |P|$ neighbors in $\cT$. The values of the parameters $n,m$ are dictated by the expansion factors $\alpha,\beta$. It is also worth pointing out that \emph{explicit} constructions of expander graphs with parameters of interest in our derivations may not be known, but their existence is guaranteed via probabilistic arguments.
We say that a set of vertices $T \in \cT$ is covered by a set $P \subseteq \cP$ if for every vertex $t \in T$, there exists a vertex $p \in P$ which is connected to $t$. We say that a vertex $t \in \cT$ is uniquely covered (or a unique neighbor) of $P$ if it is the neighbor of exactly one vertex $p \in P$. Henceforth, for a set of vertices $P$, let $N(P) \subseteq \cT$ denote the neighbors of $P$ and let $N_u(P) \subseteq \cT$ denote the set of unique neighbors of $P$. Furthermore, we say that a vertex $t \in \cT$ is covered $h$ times by $P$ if it is connected to exactly $h$ different vertices in $P$. The next results may be obtained through a straightforward modification of existing results. 

\begin{lemma}\label{lem:expp}~\cite{INR10} Suppose that $\cG$ is an $(\alpha, \beta)$-expander where every vertex in $\cP$ has $k$ neighbors and $\beta > \frac{3k}{4}$. Let $I \subseteq \cP$ be a subset of size at most $|I| \leq d$. Then for any $P \subseteq \cP$ such that $P \cap I = \emptyset$, $|P| \geq |I| + 2$ and $|P \cup I| \leq \alpha |\cP|$, we have:
\begin{align*}
\Big | N_u(P \cup I) \setminus N(I) \Big | \geq k.
\end{align*}
\end{lemma}

Thus, given the previous lemma, it follows that there exists at least one test in $N(P \cup I)$ that is not covered by an element of $I$. Using this observation, we construct the $m \times n$ binary matrix ${\bf B}$ as follows. Suppose that $\cG$ is an expander as previously described. We assume that the vertices in $\cP$ and $\cT$ are lexicographically ordered so that we can refer to the $i^{\text{th}}$ vertex in $\cT$ as $i$ and the $j^{\text{th}}$ vertex in $\cP$ as $j$. Then, for $i \in \cT$ and $j \in \cP$, 
\begin{align}\label{eq:Mit}
{\bf B} (i,j)= \begin{cases}
1, & \text{ if  an edge exists between $i$ and $j$ in $\cG$,}\\
0, & \text{ otherwise.}
\end{cases}
\end{align}
Thus, as a result of the construction for ${\bf B}$, we see that we can uniquely associate each column of ${\bf B}$ with a vertex in $\cP$ and each row of ${\bf B}$ with a vertex in $\cT$.

The next two results follow immediately from the previous discussion.

\begin{corollary}\label{cor:li} Suppose we are given two sets $I, L \subseteq \cP$ such that $I \subseteq L$. If $L$ is consistent with $I$ under ${\bf B}$ and $|L| \leq \alpha |\cP|$, then
$$ |L| < 2|I| + 2.$$
\end{corollary}

\begin{lemma}\label{lem:correctb} Suppose $\bf B$ is as defined in (\ref{eq:Mit}) and the set of infected individuals satisfies $|I| \leq d$. Then, testing with $\bf B$ recovers a set $L \subseteq \cP$ such that $|L| = \cO(d)$ and $I \subseteq L$.
\end{lemma} 

The following lemma is known~\cite{DH06} and follows from a standard randomized construction:

\begin{lemma} \label{lem:expander} Suppose that $\alpha=\frac{2d+2}{n}$ and let $m = 8 e^2 k \alpha n $, where $e$ is the base of the natural logarithm. Then, for $k = \cO(\log \frac{1}{\alpha})$ there exists an $(\alpha, \beta)$-expander graph $\cG$ with bipartition $\cP, \cT$ such that $|\cP| = n$ and $|\cT| = m$, and $\beta=\frac{3k}{4}$.
\end{lemma}

The previous result implies the following theorem.

\begin{theorem}\label{th:twob} There exists a conventional two-stage GT scheme that requires at most 
$$ \cO \Big ( 8 e^2 d \log \frac{n}{d} \Big )$$
tests and can identify a set of infected individuals of size at most $d$ from a population of size $n$.
\end{theorem}

We remark that the best known explicit constructions of bipartite expanders are still inferior to
the optimal bounds achieved by random expanders in Lemma~\ref{lem:expander}. 
For example, using \cite{ref:GUV09} one can get $O(d^{1+\alpha} (\log n)^{O(1/\alpha)})$ tests for any fixed $\alpha>0$, and \cite{ref:CRVW02}
would achieve $O(d \exp((\log \log n)^3))$ tests, similar to the derivation in
\cite{C09}.

We now discuss how to use the matrix ${\bf B}$ to design a specialized two-round SQGT testing scheme for $\tau>2$.


We focus on a special case of uniform SQGT \emph{with saturation}~\cite{acdc} for which we are given $\tau$ thresholds. The test outcome vector for a set $I$ of defectives is such that $\bs_{I}(l)=0$ if the $l^{\text{th}}$ test includes no defectives, $\bs_{I}(l)=1$ if the $l^{\text{th}}$ test includes $1$ defective, $\ldots$ , $\bs_{I}(l)=\tau-2$ if the $l^{\text{th}}$ test includes $\tau-2$ defectives and $\bs_{I}(l)=\tau-1$ if the number of defectives in the $l^{\text{th}}$ test exceeds $\tau-2$. To simplify the notation, we assume that $\tau = \left( 4\gamma\right)^{\gamma},$ for some positive integer $\gamma$. 

We show the existence of a two-round testing scheme that differs from the information theoretic lower bound from~\cite{acdc} by only a factor of roughly $\log \tau$. 
As discussed earlier, we only focus on the first round, since the second one is straightforward. The key idea used to construct the test matrix for the first round is to start with list-disjunct expander-based binary test matrix and then merge the rows via specialized linear combinations to reduce the number of tests and increase the size of the alphabet used for the codebook.

We start by introducing two matrices ${\bf S}^{(1)}$ and ${\bf S}^{(2)}$ that will be subsequently concatenated into the ``global'' SQGT matrix ${\bf S} = \begin{bmatrix} {\bf S^{(1)}} \\ {\bf S^{(2)}} \end{bmatrix}$.
Let ${\bf B}$ be as defined in (\ref{eq:Mit}) and for simplicity, assume that $\gamma\, | \,m$. Then, for $i \in [1,\frac{m}{\gamma}]$ and $j \in [1,n]$, we set
\begin{align}
{\bf S}^{(1)}(i,j) &= {\bf B}((i-1)\gamma + 1, j) + (4\gamma) {\bf B}((i-1)\gamma + 2, j) \label{eq:Mqb} \\ 
&+ (4\gamma)^2 {\bf B}((i-1)\gamma + 3, j) \notag
+ \cdots +(4 \gamma)^{\gamma-1} {\bf B}(i \gamma, j);  \\ 
{\bf S}^{(2)}(i,j) &= {\bf B}((i-1)\gamma + 1, j) +  {\bf B}((i-1)\gamma + 2, j)  \label{eq:Oqb} \\
&+ {\bf B}((i-1)\gamma + 3, j) + \cdots + {\bf B}(i \gamma, j). \notag
\end{align}
Note that both ${\bf S}^{(1)}$ and ${\bf S}^{(2)}$ are obtained linear combination of rows of {\bf B}, but the scaling factors are different. The SQGT test matrix {\bf S} has $2\frac{m}{\gamma}$ rows and consequently the same number of tests. The tests involve taking an integer number of sample units dictated by the nonbinary entries in {\bf S}. The nonbinary (semi-quantitative) test outcome vector will be denoted by $\bs$.

Let $E(a)$ denote the $(4\gamma)$-ary expansion of the natural number $a$ in vector form. More precisely, if $a = a_0 + a_1 4\gamma + a_2 \left( 4 \gamma \right)^2 + \cdots + a_{\gamma-1} \left( 4 \gamma \right)^{\gamma-1}$, then $E( a ) = \Big( a_0, a_1, \ldots, a_{\gamma-1}  \Big)$, where $a_i,i \in [0,4\gamma-1]$. Our decoding procedure operates as follows. Suppose that $\bs^{(1)} = (s^{(1)}_1, \ldots, s^{(1)}_{\frac{m}{\gamma}})$ represents the results of the (quantized) testing using the matrix (\ref{eq:Mqb}). We apply the map $E$ to $\bs^{(1)}$ entrywise. We then use an expander-based decoding procedure on this vector to recover a ``noisy'' set of test values - the ``noise'' is due to the that the matrix ${\bf S}^{(2)}(i,j)$ can handle only up to $4\gamma$ defectives. 

To this end, let $\bs' = \Big (E(s^{(1)}_1), E(s^{(1)}_2), \ldots, E(s^{(1)}_{\frac{m}{\gamma}}) \Big )=(s'_1, s'_2, \ldots, s'_{m})$ and let $\hat{\bt}^{(b)} = \Big( \lceil \frac{s'_1}{\tau} \rceil, \ldots,  \lceil \frac{s'_m}{\tau} \rceil \Big)=\Big( \hat{t}^{(b)}_1, \hat{t}^{(b)}_2, \ldots, \hat{t}^{(b)}_m \Big) \in \{0,1\}^m$. Note that $\hat{\bt}_i^{(b)}=0$ if $s_i'>0$ and zero otherwise. For shorthand, we write
$f_{\tau \to b}\Big( \bs^{(1)} \Big) = \hat{\bt}^{(b)}.$
We have the following claim.

\begin{claim}\label{cl:hamm} Let $\bt \in \{{0,1\}}^m$ denote the test output based on the binary matrix ${\bf B}$, let ${\bs}^{(1)}$ be the test output generated via ${\bf S}^{(1)}$ and let $\hat{\bt}^{(b)}$ be as defined above.
Then, 
$$ d_H\Big ( \hat{\bt}^{(b)}, \bt \Big) \leq \frac{d k}{4}.$$
\end{claim}
\begin{IEEEproof} Let $f_{\tau \to b}(s_i^{(1)}) = (t_{ \gamma (i-1) +1}, \ldots, t_{ \gamma i} )$ be the mapping that corresponds to ${\bs}_i^{(1)}.$ For some $j \in [0,\gamma-1]$, let vertex $(i-1)\gamma + j \in \cT$ be covered $ \geq 4 \gamma$ times. Such a vertex may be in error (due to the use of the $4 \gamma$-ary expansion). Since the set $I \subseteq \cP$ has at most $|I| k$ neighbors in $\cT \subseteq \cG$, it follows from an averaging argument that 
$$ \Big| \Big \{ (i,j) : \text{vertex $(i-1)\gamma + j \in \cT$ is covered $\geq 4 \gamma$ times} \Big \} \Big| $$
$$\leq \frac{|I| k}{4 \gamma }. $$
Let $(i-1)\gamma + \ell \in \cT$ be a vertex in $\cT$ which is covered at least $4 \gamma$ times (if no such vertex exists, we are error-free and do not have to prove anything further). In this case we may have $f_{\tau \to b} \Big( s^{(1)}_{(i-1)\gamma + \ell} \Big) = \Big( \hat{t}^{(b)}_{(i-1)\gamma + 1}, \hat{t}^{(b)}_{(i-1)\gamma + 2}, \ldots, \hat{t}^{(b)}_\gamma \Big) \neq \Big( t_{(i-1)\gamma + 1}, t_{(i-1)\gamma + 2}, \ldots, t_\gamma \Big)$; 
in the worst case $\hat{t}^{(b)}_{(i-1)\gamma + 1} \neq$ $t_{(i-1)\gamma + 1},$ $\ldots,$ $\hat{t}^{(b)}_{i\gamma} \neq t_{i \gamma}$. This implies that for every $(i,\ell)$ there are at most $\gamma$ instances where $\hat{t}^{(b)}_{v} \neq t_{v}$, which gives the desired result.
\end{IEEEproof}


As a result of the previous lemma, it follows that we can recover a binary vector $\hat{\bt}^{(b)}$ that is within Hamming distance $\frac{dk}{4}$ of the binary test result $\bt$ based on $\bf{B}$. Thus, we have to recover the set of infected individuals given a noisy set of test outcomes. To correct errors, we make use of the test outcome generated by the matrix ${\bf S}^{(2)}$; this matrix renders the errors in $\bt$ ``asymmetric,'' which simplifies the problem. Here, the term ``asymmetric'' refers to the fact to be addressed in Claim 7 that $\hat{\bft} \geq \bft$ so that in $\bft$ a $0$ can change to a $1$ but not otherwise. More precisely, we use ${\bf S}^{(2)}$ to identify tests in ${\bf S}^{(1)}$ that contain $>4 \gamma$ defectives. Note that if at least $4 \gamma$ infected individuals are present in some test pool $i$, then the entries indexed by $(i-1)\gamma + 1, (i-1)\gamma + 2, \ldots, i\gamma$ of $\hat{\bt}^{(b)}$ may be in error. 

Let ${\bs}^{(2)} = (s^{(2)}_1, \ldots, s^{(2)}_{m/ \gamma}) \in [0,\tau-1]^{\frac{m}{\gamma}}$ be the test outcomes of ${\bf S}^{(2)}$. Define a vector
\begin{align}\label{eq:bot}
\overline{t}^{(b)}_j =\begin{cases} 1, &\text{ if } s^{(2)}_{\lceil \frac{j}{m} \rceil} \geq 4\gamma, \\
\hat{t}^{(b)}_j, &\text{ otherwise.}
\end{cases}
\end{align}
Similarly as before, for ${\bs} = ({\bf s}^{(1)}, \bfs^{(2)})$ we write 
$$ \overline{f}_{\tau \to b}({\bf s}) = \overline{\bt}^{(b)}.$$

The following straightforward claim follows from the previous discussion and the observations in Claim~\ref{cl:hamm}.

\begin{claim}\label{cl:odiff} Let $\overline{\bt}^{(b)} = \overline{f}_{\tau \to b}({\bs})$. Then, $\overline{\bt}^{(b)} \geq {\bt}$, and 
$$ d_H(\overline{\bt}^{(b)}, {\bt}) \leq \frac{d k}{4}. $$
\end{claim}
We next generate a list $L$ of potentially infected individuals consistent with the outcome of the tests $\overline{\bt}^{(b)}$. The next lemma, which uses the same ideas as Lemma~\ref{lem:expp}, describes an upper bound on the size of $L$. 

\begin{lemma}\label{ub:list} Suppose that $\bs \in [0,\tau-1]^{2\frac{m}{\gamma}}$ is the result of the tests in (\ref{eq:Mqb}) and (\ref{eq:Oqb}) and $\overline{\bt}^{(b)} = \overline{f}_{\tau \to b}(\bs)$. Then the size of any list of defectives from $\cP$ consistent with $\overline{\bt}^{(b)} = f_{\tau \to b}(\bfs)$ is at most $\cO(d)$.
\end{lemma}
\begin{IEEEproof} Recall that in our setup the graph $\cG$, which is used to construct ${\bf B}$ and also ${\bf S}^{(1)}$, is an $(\alpha, \beta)$-expander. Hence, every vertex in $\cP$ has $k$ neighbors and $\beta > \frac{3k}{4}$. As before, let $I \subseteq \cP$ denote the set of infected individuals such that $|I| \leq d$. Let $\bt_{I} \in \{0,1\}^m$ be the output of the tests dictated by ${\bf B}$. We show that given a $S \subseteq \cP$ such that $S \cap I = \emptyset$ and $|S| \geq \cO(d)$, $S \cup I$ cannot be consistent with $\overline{\bt}^{(b)}$ under ${\bf B}$.

Let $S' = S \cup I \subseteq \cP$. Using the same arguments as in the proof of Lemma~\ref{lem:expp}, we can show that the number of unique neighbors of $S'$ satisfies
$$ N_u(S') \geq \frac{k |S|}{2}.$$
Let $E = \{ j :  \overline{t}_j^{(b)} > t_j \}$. Since $N(I) \leq d k$ and $| E| \leq \frac{d k}{4}$ from Claim~\ref{cl:odiff}, it follows that 
\begin{align*}
\Big | N_u(S') \setminus (I \cup E) \Big| \geq \frac{k |S|}{2} - ( dk + \frac{dk}{4} ),
\end{align*}
which implies that if $|S| > \frac{10d}{8}$, then there exists a unique neighbor of $S'$ which is not in error and is also not already covered by an element in $I$. This implies that $N(S')$ is not consistent with $\overline{\bt}^{(b)}$.
\end{IEEEproof}

The following theorem follows from the previous discussion and from Claim~\ref{cl:hamm} and Lemma~\ref{ub:list}.

\begin{theorem} There exists a nonbinary two-stage GT scheme that given $\tau = (4\gamma)^{\gamma}$ thresholds and 
$$ \cO \Big ( \frac{8 e^2 d}{\gamma} \log \frac{n}{d} \Big )$$
tests that can identify a set of infected individuals of size at most $d$ in a population of size $n$.
\end{theorem}
\begin{IEEEproof} We prove the result by describing a simple method for recovering a set $L$ of size $\cO(d)$ which contains the set of defectives $I$. First, we generate the vector $\overline{\bt}^{(b)} = \overline{f}_{\tau \to b}({\bs})$ from our non-binary test outcomes. We initialize $L = \emptyset$. Then, for every $p \in \cP$, if $L \cup \cP$ is consistent with $\overline{\bt}^{(b)}$, we update $L = L \cup \{{p\}}$. Otherwise, we do not change $L$. At the end of this process we have $I \subseteq L$. Furthermore, according to Lemma~\ref{ub:list}, $|L| \leq \cO(d)$. The result now follows from Theorem~\ref{th:twob}.
\end{IEEEproof}

\section{Nonadaptive SQGT} \label{sec:oneround}
We describe next constructive nonadaptive testing schemes, which in the asymptotic regime require at most $\cO ( \frac{d^2}{\gamma} \log n  )$ tests, with $\tau = (4\gamma)^\gamma$. 
Our approach builds upon the construction by Porat and Rothschild (PR construction)~\cite{pr11}, which makes use of non-binary error-correcting codes. Our key result is described in Lemma~\ref{lem:dist}.   

Let $\cC \in \mathbb{F}_q^{m/q}$ be a $q$-ary linear error-correcting code, where $q$ is an odd prime, of minimum distance $\delta \frac{m}{q}$, $q=\cO(d)$, and dimension $\log_q(n)$. The PR construction works by uniquely associating each individual in the population of size $n$ with a codeword in $\cC$. Under this setup, the test matrix ${\bf B}^{(PR)} = (b_{(c,x),j})_{c \in [1,\frac{m}{q}], x \in [0,q-1], j \in [1,n]}$ is defined as 
\begin{align*}
b_{(c,x),j} = \begin{cases}
&1, \text{ if } \bx^{(j)}_c = x,\\
&0, \text{ otherwise,}
\end{cases}
\end{align*}
where $\bx^{(j)}$ is the $j$-th codeword of $\cC$. In words, the test indexed by $(c,v)$ contains the codewords (individuals) from $\cC$ whose $c$-th coordinate equals $x$. 

Our approach for designing a nonadaptive testing scheme is similar to that for the adaptive setting. Each test can be generated by taking a linear combination of $\gamma$ rows of ${\bf B}^{(PR)}$. 
The total number of tests equals $2 \frac{m}{q} \lceil \frac{q}{\gamma} \rceil = \cO(\frac{m}{\gamma} )$, and once again the tests are represented by ${\bf S} = \begin{bmatrix} {\bf S}^{(1)} \\ {\bf S}^{(2)} \end{bmatrix},$ where ${\bf S}^{(2)} = \Big(s^{(2)}_{(c,r),j} \Big)_{c \in \frac{m}{q}, r \in [0, \lceil \frac{q}{\gamma} \rceil -1], j \in [1,n]},$ is defined as follows:
\begin{align*}
s^{(2)}_{(c,r),j} = \begin{cases}
&1, \text{ if } \bx^{(j)}_c = [(r-1)\gamma, \min\{ r \gamma - 1, q-1\} ],\\
&0, \text{ otherwise.}
\end{cases}
\end{align*}
In words, the test in ${\bf S}^{(2)}$ indexed by $(c,r)$ contains the codewords (individuals) from $\cC$ whose $c$-th coordinate has a value between $(r-1)\gamma$ and the minimum of $r\gamma -1, q-1$. Note that the reason for using the minimum in the previous range of values is a consequence of the fact that we assumed $q$ to be an odd prime. The tests in ${\bf S}^{(1)}$ are defined similarly: Suppose that $x^{(j)}_c = (r-1) \gamma + v'$ where $v' \in [0,\gamma-1]$. Then,
\begin{align*}
s^{(1)}_{(c, r),j} = (4 \gamma)^{v'}.
\end{align*}
For shorthand, we refer to the codewords in the $(c,r)$-th test in ${\bf S}^{(1)}$ as $T_{(c,r)} \subseteq \cC$. 

\begin{claim}\label{cl:radix} Suppose that the number of infected individuals in the test indexed by $(c,r)$ is at most  $4\gamma-1$ so that 
$$\Big |T_{(c,r)} \cap I \Big | \leq 4\gamma -1.$$ 
Then, given the output of the test $T_{(c,r)}$ we can uniquely determine     
$$ \Big | \big \{ \bfx \in I : x_c = x \big \} \Big |,$$
for $x \in [(r-1)\gamma, r\gamma -1]$.
\end{claim}

Let $\bf{1}^n$ denote the all-ones vector of length $n$. We assume that our code $\cC$ is such that $\bf{1}^n \in \cC$. Henceforth, let
\begin{align}\label{eq:overD}
\overline{I} = \Big \{ {\bf y} + i \cdot \textbf{1}^n : {\bf y} \in I \Big\}
\end{align}
for all $i \in \{-\gamma + 1, \ldots, -1, 0, 1, \ldots, \gamma-1 \}$.

\begin{claim}\label{cl:card} Let $\bx \in \cC \setminus I$ be such that $\bx \in T_{(c,r)}$. Suppose that for an integer $\ell$ we have
\begin{align*}
\Big | \big \{ {\bf z} \in \overline{I} : x_c = z_c  \big \} \Big | \leq \ell.
\end{align*}
Then, 
\begin{align*}
\Big | T_{(c,r)} \cap I \Big | \leq \ell.
\end{align*}
\end{claim}
\begin{IEEEproof}  This follows since if $\bx \in T_{(c,r)}$, then $x_c = r (\gamma-1) + v'$ for some $v' \in [0,\gamma-1]$. If ${\bf z} \in T_{(c,r)} \cap I$, then $z_c = r(\gamma-1) + v''$ for some $v'' \in [0,\gamma-1]$. Since $v', v'' \in [0,\gamma-1]$, it follows that $z_c + (v'-v'') = r (\gamma-1) + v' = x_c$ where $(v' - v'') \in \{ -\gamma+1, \ldots, -1,0,1, \ldots, \gamma -1\}$. This in turn implies that $z_c + (v'-v'')$ is the value of component $c$ of a vector from the set $\overline{I}$.
\end{IEEEproof}

We also need the following result. 
\begin{claim}\label{cl:confcri} Suppose that $\bx \in \cC \setminus I$ is such that $\bx \in T_{(c,r)}$. If there exists an index $c \in [n]$ satisfying
\begin{align}
\Big | \big \{ \bz \in \overline{I} : x_c = z_c \big \} \Big | \leq 4 \gamma -1, \label{eq:overcl}
\end{align}
and
\begin{align}
\Big | \big \{ \bz \in {I} : x_c = z_c \big \} \Big | =0, \label{eq:regcl}
\end{align}
then given the output of the tests dictated by ${\bf S}^{(1)}$,$ {\bf S}^{(2)}$ we can determine that $\bx \notin I$.
\end{claim}
\begin{IEEEproof} From Claim~\ref{cl:card} and if (\ref{eq:overcl}) holds, we have that $\Big | \big \{ I \cap T_{(c,r)}  \big \} \Big | \leq 4\gamma-1$. Then from Claim~\ref{cl:radix}, since the number of infected individuals in $T_{(c,r)}$ is at most $4\gamma-1$, we have $ \Big | \big \{ \bz \in I : z_c = x_c \big \} \Big |=0$ using the test outputs of $T_{(c,r)}$. 
\end{IEEEproof}

\begin{lemma}\label{lem:dist} If $\cC$ has minimum distance $\delta > 1 - \frac{1}{2 d}$, the tests ${\bf S}$ uniquely determine the set $I$ of defectives.
\end{lemma}
\begin{IEEEproof} According to Claim~\ref{cl:confcri}, we need to show that (\ref{eq:overcl}) and (\ref{eq:regcl}) hold for any $\bx \in \cC \setminus I$. We start by showing that (\ref{eq:regcl}) holds. In particular, we show a stronger claim that there exists a set $C^{(1)} \subseteq \frac{m}{q}$ of size at least $\frac{m}{2q} + 1$ where for any $c \in C^{(1)}$, we have
\begin{align}
x_c \neq y_c, \label{eq:xpdp}
\end{align}
where $\by = (y_1, \ldots, y_{\frac{m}{q}}) \in I$. Note that this implies that the number of coordinates of $\bx$ which agree in value with an element of $I$ is at most $\frac{m/q}{2} - 1$. Since any two elements in $\cC$ can agree in at most $(1-\delta)\frac{m}{q}$ coordinates and $\delta > 1- \frac{1}{2d}$, it follows that 
\begin{align*}
\Big | \big \{ c : x_c = y_c, {\bf y} \in I \big \} \Big | \leq d(1-\delta) \frac{m}{q} < \frac{m}{2q}.
\end{align*}

Next, we show that for at least one coordinate in $C^{(1)}$, (\ref{eq:overcl}) holds as well. First, note that
\begin{align*}
\Big| \Big \{ \big( \by,c \big) \in \overline{I} \times \frac{m}{q} : x_c = y_c \Big \} \Big | \leq 2\gamma d (1-\delta) \frac{m}{q} <  \gamma \frac{m}{q},
\end{align*}
so that for a randomly chosen coordinate $c \in \frac{m}{q}$, 
\begin{align*}
E \Big[ \Big| \big \{ \by \in \overline{I} : x_c = y_c \}  \Big | \Big ] < \gamma.
\end{align*}
Invoking Markov's inequality we get
\begin{align*}
\text{Pr} \Big(\big| \big \{ {\bf y} \in \overline{I} : x_c = y_c \}  \big | \geq 4 \gamma \Big) < \frac{1}{4}.
\end{align*}
Therefore, it follows that there exists a set of coordinates $C^{(2)} \subseteq \frac{m}{q}$ of size at least $\frac{m}{2q}$ such that for any $c \in C^{(2)}$ 
\begin{align*}
\Big| \Big \{ {\bf y} \in \overline{I} : x_c = y_c  \Big \} \Big | < 4 \gamma.
\end{align*}
Since $|C^{(2)}| \geq \frac{m}{2q}$ and $C^{(1)} \geq \frac{m}{2q} + 1$, it follows that $|C^{(1)} \cap C^{(2)}| \geq 1$. Letting $c^{*} \in C^{(1)} \cap C^{(2)}$ we have $\Big | \big \{ {\bf y} \in \overline{I} : x_{c^{*}} = y_{c^{*}} \big \} \Big | \leq 4\gamma -1$ and $\Big | \big \{ {\bf y} \in {I} : x_{c^{*}} = y_{c^{*}} \big \} \Big | =0$. By Claim~\ref{cl:confcri}, we conclude that $\bx \not \in I$.
\end{IEEEproof}

\textbf{Open Problems.} Despite only a small gap remaining between the lower bound and the actual constructions for the 
saturation model, many other problems remain open and include:
\begin{itemize}
\item Extending the nonadaptive and two-round constructions for general quantization thresholds under the SQGT model;
\item Deriving bounds and test strategies for consecutive defective models~\cite{bui2021improved,colbourn1999group}, as these capture the order of arrivals into testing queues;
\item Addressing generalized binomial SQGT algorithms~\cite{hwang1975generalized}.
\end{itemize}

\balance
\bibliographystyle{IEEEtran}
\bibliography{bibliography}

\begin{thebibliography}{10}
\providecommand{\url}[1]{#1}
\csname url@samestyle\endcsname
\providecommand{\newblock}{\relax}
\providecommand{\bibinfo}[2]{#2}
\providecommand{\BIBentrySTDinterwordspacing}{\spaceskip=0pt\relax}
\providecommand{\BIBentryALTinterwordstretchfactor}{4}
\providecommand{\BIBentryALTinterwordspacing}{\spaceskip=\fontdimen2\font plus
\BIBentryALTinterwordstretchfactor\fontdimen3\font minus
  \fontdimen4\font\relax}
\providecommand{\BIBforeignlanguage}[2]{{%
\expandafter\ifx\csname l@#1\endcsname\relax
\typeout{** WARNING: IEEEtran.bst: No hyphenation pattern has been}%
\typeout{** loaded for the language `#1'. Using the pattern for}%
\typeout{** the default language instead.}%
\else
\language=\csname l@#1\endcsname
\fi
#2}}
\providecommand{\BIBdecl}{\relax}
\BIBdecl

\bibitem{D43}
R.~Dorfman, ``The detection of defective members of large populations,''
  \emph{Annals of Mathematical Statistics}, vol.~14, pp. 436--440, 1943.

\bibitem{KS64}
W.~Kautz and R.~Singleton, ``Nonrandom binary superimposed codes,'' \emph{IEEE
  Transactions on Information Theory}, vol.~10, pp. 363--377, 1964.

\bibitem{DH06}
D.-Z. Du and F.-K. Hwang, \emph{Pooling Designs and Nonadaptive Group
  Testing}.\hskip 1em plus 0.5em minus 0.4em\relax World Scientific, 2006.

\bibitem{kairouz}
H.~A. Inan, P.~Kairouz, M.~Wootters, and A.~{\"O}zg{\"u}r, ``On the optimality
  of the {Kautz-Singleton} construction in probabilistic group testing,''
  \emph{IEEE Transactions on Information Theory}, vol.~65, no.~9, pp.
  5592--5603, 2019.

\bibitem{W85}
J.~Wolf, ``Born again group testing: Multiaccess communications,'' \emph{IEEE
  Transactions on Information Theory}, vol.~31, no.~2, pp. 185--191, 1985.

\bibitem{D04}
A.~Dyachkov, ``Lectures on designing screening experiments,'' 2004, lecture
  Note Series 10.

\bibitem{D06}
P.~Damaschke, ``Threshold group testing,'' in \emph{General Theory of
  Information Transfer and Combinatorics}, ser. Lecture Notes in Computer
  Science, vol. 4123, 2006, pp. 707--718.

\bibitem{L75}
B.~Lindstrom, ``Determining subsets by unramified experiments,'' \emph{A Survey
  of Statistical Design and Linear Models}, 1975.

\bibitem{DH00}
D.-Z. Du and F.~Hwang, \emph{Combinatorial Group Testing and its Applications},
  2nd~ed.\hskip 1em plus 0.5em minus 0.4em\relax World Scientific, 2000.

\bibitem{AM110}
A.~{Emad}, J.~{Shen}, and O.~{Milenkovic}, ``Symmetric group testing and
  superimposed codes,'' in \emph{2011 IEEE Information Theory Workshop}, 2011,
  pp. 20--24.

\bibitem{EO14}
A.~{Emad} and O.~{Milenkovic}, ``Semiquantitative group testing,'' \emph{IEEE
  Transactions on Information Theory}, vol.~60, no.~8, pp. 4614--4636, 2014.

\bibitem{EO16}
------, ``Code construction and decoding algorithms for semi-quantitative group
  testing with nonuniform thresholds,'' \emph{IEEE Transactions on Information
  Theory}, vol.~62, no.~4, pp. 1674--1687, 2016.

\bibitem{DR81}
A.~G. D'yachkov and V.~V. Rykov, ``A coding model for a multiple-access adder
  channel,'' \emph{Probl.\ Perdachi Inform.{\ }}, pp. 26--32, 1981, in Russian.

\bibitem{DR83}
A.~Dyachkov and V.~Rykov, ``A survey of superimposed code theory,''
  \emph{Problems of Control and Information Theory}, vol.~12, no.~4, pp.
  229--242, 1983.

\bibitem{acdc}
R.~Gabrys, S.~Pattabiraman, V.~Rana, J.~{a}o Ribeiro, M.~Cheraghchi,
  V.~Guruswami, and O.~Milenkovic, ``{AC-DC}: Amplification curve diagnostics
  for {Covid-19} group testing,'' 2020, arXiv:2011.05223.

\bibitem{gbgt75}
F.~Hwang, ``A generalized binomial group testing problem,'' \emph{Journal of
  the American Statistical Association}, vol.~70, no. 352, pp. 923--926, 1975.

\bibitem{list-disjunct}
H.~Q. Ngo, E.~Porat, and A.~Rudra, ``Efficiently decodable error-correcting
  list disjunct matrices and applications,'' in \emph{International Colloquium
  on Automata, Languages, and Programming}.\hskip 1em plus 0.5em minus
  0.4em\relax Springer, 2011, pp. 557--568.

\bibitem{pr11}
E.~Porat and A.~Rothschild, ``Explicit nonadaptive combinatorial group testing
  schemes,'' \emph{IEEE Transactions on Information Theory}, vol.~57, no.~12,
  pp. 7982--7989, 2011.

\bibitem{INR10}
P.~Indyk, H.~Q. Ngo, and A.~Rudra, ``Efficiently decodable non-adaptive group
  testing,'' in \emph{Proceedings of the twenty-first annual ACM-SIAM symposium
  on Discrete Algorithms}.\hskip 1em plus 0.5em minus 0.4em\relax SIAM, 2010,
  pp. 1126--1142.

\bibitem{C09}
M.~Cheraghchi, ``Noise-resilient group testing: Limitations and
  constructions,'' \emph{Discrete Applied Mathematics}, vol. 161, no.~1, pp.
  81--95, 2013, preliminary version in Proceedings of the {FCT 2009}. arXiv
  manuscript published in 2008.

\bibitem{BGV05}
A.~De~Bonis, L.~Gasieniec, and U.~Vaccaro, ``Optimal two-stage algorithms for
  group testing problems,'' \emph{SIAM Journal on Computing}, vol.~34, no.~5,
  pp. 1253--1270, 2005.

\bibitem{ref:CN20}
M.~Cheraghchi and V.~Nakos, ``Combinatorial group testing schemes with
  near-optimal decoding time,'' in \emph{Proceedings of the 61st Annual {IEEE}
  Symposium on Foundations of Computer Science {(FOCS)}}, 2020.

\bibitem{F96}
Z.~F{\"u}redi, ``On $r$-cover-free families,'' \emph{Journal of Combinatorial
  Theory, Series A}, vol.~73, no.~1, pp. 172--173, 1996.

\bibitem{ref:GUV09}
V.~Guruswami, C.~Umans, and S.~Vadhan, ``Unbalanced expanders and randomness
  extractors from {Parvaresh-Vardy} codes,'' \emph{Journal of the ACM},
  vol.~56, no.~4, 2009.

\bibitem{ref:CRVW02}
M.~Capalbo, O.~Reingold, S.~Vadhan, and A.~Wigderson, ``Randomness conductors
  and constant-degree expansion beyond the degree/2 barrier,'' in
  \emph{Proceedings of the $34$th Annual {ACM} Symposium on Theory of Computing
  ({STOC})}, 2002, pp. 659--668.

\bibitem{bui2021improved}
T.~V. Bui, M.~Cheraghchi, and T.~D. Nguyen, ``Improved algorithms for
  non-adaptive group testing with consecutive positives,'' \emph{arXiv preprint
  arXiv:2101.11294}, 2021.

\bibitem{colbourn1999group}
C.~J. Colbourn, ``Group testing for consecutive positives,'' \emph{Annals of
  Combinatorics}, vol.~3, no.~1, pp. 37--41, 1999.

\bibitem{hwang1975generalized}
F.~Hwang, ``A generalized binomial group testing problem,'' \emph{Journal of
  the American Statistical Association}, vol.~70, no. 352, pp. 923--926, 1975.

\end{thebibliography}

\end{document}